\documentclass[a4paper,11pt]{article}
\pdfoutput=1

\usepackage{jheppub}

\usepackage[T1]{fontenc}

\usepackage{graphicx}
\usepackage{enumitem}
\usepackage{latexsym}
\usepackage{amsfonts}
\usepackage{amssymb}
\usepackage[dvipsnames]{xcolor}
\usepackage{amsmath}
\usepackage{slashed}
\usepackage{dcolumn}
\usepackage{verbatim}
\usepackage{float}
\usepackage{multirow}
\usepackage{xspace}
\usepackage[normalem]{ulem}
\usepackage[caption=false]{subfig}
\usepackage{bm}
\usepackage{bbm}

\renewcommand{\tilde}{\widetilde} 

\newcommand{\beq}{\begin{equation}}
\newcommand{\eeq}{\end{equation}}
\newcommand{\bea}{\begin{eqnarray}}
\newcommand{\eea}{\end{eqnarray}}

 \def\d{{\rm d}}

\newcommand{\refjnl}[1]{{\rm#1}}

\def\apj{\refjnl{ApJ}}                 
\def\apjs{\refjnl{ApJS}}               
\def\aap{\refjnl{A\&A}}                

\def\mnras{\refjnl{MNRAS}}             
\def\prd{\refjnl{Phys.~Rev.~D}}        
\newcommand{\vtwo}[1]{{#1}}

\renewcommand{\paragraph}[1]{~\\ \noindent{\bf \emph{#1} --}}

\newcommand{\MeV}{\text{MeV}}
\newcommand{\GeV}{\text{GeV}}

\title{Cosmic microwave background constraints on extended dark matter objects}
\author[a]{Djuna Croon,}
\author[a]{Sergio Sevillano Mu\~noz}

\affiliation[a]{Institute for Particle Physics Phenomenology, Department of Physics, Durham University, \\Durham DH1 3LE, U.K.
}

\emailAdd{djuna.l.croon@durham.ac.uk}
\emailAdd{sergio.sevillano-munoz@durham.ac.uk}
\preprint{IPPP/24/11}

\abstract{
Primordially formed extended dark objects would accrete baryonic matter and impact the ionisation history of the Universe. Insisting on consistency with the anisotropies of the cosmic microwave background, we derive constraints on the dark matter fraction for various classes of objects, of different sizes. We introduce a novel scaling technique to speed up numerical calculations and  release our calculation framework in the form of a \href{https://gitlab.com/SergioSevillano/edo-accretion}{Mathematica notebook}. Conservatively, we focus on spherical accretion and collisional ionisation. We find strong constraints limiting the dark matter fraction to subpercent level for objects of up to $10^4$ AU in size.
}
\begin{document}
\maketitle
\flushbottom

\section{Introduction}
Despite the fact that dark matter (DM) comprises $\sim 85\%$ of the total amount of matter in our Universe, very little is known about its mass: the potential mass range of DM candidates spans over 80 orders of magnitude. At the high end of this spectrum, macroscopic DM candidates, with masses ranging from comparable to large asteroids to large stars, offer an alternative to particle- and wave- like DM. Such celestial objects, formed in the early universe, are so sizeable that they can in principle be detected via their gravitational effects. As such, important constraints  have been set using gravitational (micro-)lensing techniques
(e.g. \cite{Alcock_1998,Niikura:2017zjd,Zumalacarregui:2017qqd,Niikura:2019kqi,Green:2020jor,Fairbairn:2017sil,Croon:2020ouk,Bai:2020jfm,Croon:2020wpr,CrispimRomao:2024nbr,DeRocco:2023hij}), and gravitational waves (e.g. \cite{Bird:2016dcv,Ali-Haimoud:2017rtz,Kavanagh:2018ggo,Bertone:2019irm,LIGOScientific:2019kan,Chen:2019irf,Franciolini:2021nvv,Croon:2022tmr}) on these objects as DM candidates. 

Formed in the early Universe, celestial DM objects are also expected to accrete throughout their lifetime. 
Thus, one can study the interaction between accreting compact objects formed in the early Universe and the cosmic microwave background (CMB). As the objects accrete matter, they emit high-energy radiation, which can influence the ionisation and thermal state of the surrounding medium. This process of energy injection can lead to alterations in the ionisation fraction, describing the ratio of free electrons to total hydrogen atoms in the universe. Changes in the ionisation fraction prior to and during the epoch of recombination have the potential to leave discernible imprints on the CMB.

Measurements of the CMB by satellites such as Planck have yielded precise data on its temperature fluctuations and polarisation, and therefore the ionisation fraction.
This has enabled stringent constraints to be placed on the abundance of primordial black holes (PBHs), popular examples of primordial compact objects, in the very high mass range $ \rm 10 \, M_\odot < M_{\rm PBH} < 10^4 \, M_\odot$  (see e.g. \cite{carr1981pregalactic,Ricotti:2007au,Ali-Haimoud:2016mbv,Poulin:2017bwe}).
Ref.~\cite{Bai:2020jfm} generalised the analysis of PBH to objects with a uniform density profile comprising 100\% of DM. In this work it was found that even relatively dilute objects of several hundreds of AU could cause observable levels of ionisation in the CMB. 

In this work, we generalise the accretion formalism further to account for extended dark objects (EDOs) of arbitrary radius and density profiles.
In contrast to~\cite{Bai:2020jfm}, which focused on an analytical calculation of the ionisation fraction, our computations are primarily numerical which allows them to be easily adapted to any mass profile.
We introduce a scaling technique that significantly speeds up the numerical calculation. 
The primary purpose of this work is to find the constraints on the fraction of DM that the extended dark objects can comprise. We find stringent constraints for objects over $10^6\,\rm R_\odot$ or $ 10^4 \rm \, AU$ in the mass range above $10^3 \, \rm M_\odot $, complementary to those found in microlensing \cite{Fairbairn:2017sil,Croon:2020ouk,Bai:2020jfm,Croon:2020wpr} {and projections such as those found in pulsar timing delay studies \cite{Dror:2019twh}}, and dominating the constraints found by \cite{Graham:2023unf,Graham:2024hah} from the dynamical heating of stars in ultrafaint dwarfs in the high mass range.

The structure of this paper can be conveniently inferred from the Table of Contents. 

\section{Accretion}\label{sec:accretion}

The gravitational influence of extended dark matter objects can lead to the accretion of matter, leading to a possible alteration of the ionisation  history of the universe. To calculate the actual impact, first we need to obtain the density profile of accreted matter in the dark matter object, as well as how this translates into its temperature, ionisation  fraction and electron/proton number. In this section, we will describe how this is calculated for a generic set of EDO mass functions, closely following the steps in Refs.\cite{Ali-Haimoud:2016mbv,Bai:2020jfm}. 

\subsection{Baryonic matter}
As a good approximation, we can assume that {at the relevant redshifts} the universe is made up of {non-relativistic} electrons, protons and hydrogen atoms, which we refer to in this work as \textit{baryonic matter}. In this way, we can define the density of baryonic matter as
\begin{equation}
	\rho =m_e n_e +m_pn_p +m_H n_H ,
\end{equation}
where $n_e(r,t)$, $n_p(r,t)$ and $n_H(r,t)$ are the number densities of electrons, protons and hydrogen atoms, respectively, and $m_e\simeq0.511 ~\MeV$ and $m_p\approx m_H \simeq 0.938 ~\GeV$ their corresponding masses. Assuming local charge neutrality implies 

\begin{equation}
	n_e =n_p ,
\end{equation}
which allows us to quantify each parameter through the fraction of ionized hydrogen atoms, given by
\begin{equation}
	x_e =\frac{n_e }{n_e  +n_H }.
\end{equation}
Therefore, considering that $m_e\ll m_p$, we obtain
\begin{align}
	n_e(r,t)=&n_p(r,t)\approx \frac{1}{m_p} x_e(r,t) \rho(r,t),\label{eq: ne np}\\
	n_H(r,t)=&\frac{1}{m_p}(1-x_e(r,t))\rho(r,t).\label{eq: nH}
\end{align}
Having defined these parameters, now we can address the evolution and thermodynamics of the initially homogeneous baryonic fluid.
\subsubsection{Evolution and thermodynamics of baryonic matter}\label{sec: Thermo}
The dynamic evolution of the baryon fluid around an EDO\footnote{We will calculate the accretion profile for a single, isolated extended dark matter object. This approximation will be justified later in section~\ref{sec: ionisation }, when studying these objects in a cosmological context.} will be dictated by its mass density $\rho(r,t)$, fluid velocity $\vec{v}(\vec{r},t)$, and internal energy density $\mathcal{E}(r,t)$ via the set of Navier-Stokes equations
\begin{align}
\dot{\rho} & + \frac{1}{r^2}(r^2\rho v )'=0,\label{eq:N-S 1}\\
\rho &\dot{v}  + \rho  v v'+P'=\rho g,\\
\rho&\dot{(\mathcal{E}/\rho)}  +\rho v (\mathcal{E}/\rho)' +P \frac{1}{r^2}(r^2v)'=\dot{q},\label{eq: N-S 3}
\end{align}
which ensures the conservation of mass, energy, and momentum. 
Herein, we have used $\dot{}=\partial/\partial t$, $'=\partial/\partial r$, and have imposed rotational symmetry such that $\vec{v}(\vec{r},t)=v(r,t)\vec{r}/r$. In these equations, some terms are still to be defined and specified, which we proceed to do now. Firstly, in the second equation, the $g(r,t)$ term corresponds to the radial component of the potential energy sources acting on the baryonic fluid. In particular, two  terms are sourcing this component, the main one is given by the gravitational pull exerted by the EDO,
 \begin{equation}
 	\vec{g}_{\rm grav}(\vec{r})=g_{\rm grav}(r) \frac{\vec{r}}{r}, \qquad \text{with} \qquad g_{\rm grav}(r)=-G \frac{M(r)}{r^2},
 \end{equation}
 where $G$ is Newton's gravitational constant and $M(r)$ is the mass distribution of the extended dark matter object. In addition to the gravitational pull, there is also a drag term arising from the interactions between the accreting and the CMB radiation, but {as in \cite{Bai:2020jfm}} it can be ignored for the systems we will be studying in this work. 
 
In terms of the internal energy density, $\mathcal{E}(r,t)$, and pressure, $P(r,t)$, of the baryonic matter, we can use Fermi-Dirac statistics for electrons and Maxwell-Boltzman statistics for protons and hydrogen, leading to 
\begin{align}
\mathcal{E}&=\frac{3}{2} T\left[n_e f_\mathcal{E}(T/m_e)+n_p+n_H\right]\\
P&=T\left[n_e f_P(T/m_e)+n_p+n_H\right],
\end{align}	
where $T(r,t)$ is set to be the common temperature of the baryon fluid in local equilibrium. The function $f_\mathcal{E}(X)$ varies from $f_\mathcal{E}(X\ll1)\sim1$ to $f_\mathcal{E}(X\gg1)\sim7\pi^4/(270 \zeta(3))\simeq2.10$, and $f_P(X)$ from $f_P(X\ll1)\sim1$ to $f_P(X\gg1)\sim7\pi^4/(540\zeta(3))\simeq1.05$. These two functions introduce the relativistic behaviour of the electron, which changes its thermal properties when $T\gg m_e$. Notice that, for a complete analysis, we would also have to include similar terms for the proton and hydrogen atom, but throughout this work we will just consider temperatures much smaller than the mass of the proton, for which the equivalent functions are set to unity. Using the definitions in Eqs.~\eqref{eq: ne np} and \eqref{eq: nH} we can simplify these two parameters as follows
\begin{align}
	\mathcal{E}&=\frac{3}{2} \frac{1}{m_p}T\rho\left[1+ x_e f_\mathcal{E}(T/m_e)\right]\\
	P&=\frac{1}{m_p}T\rho\left[1+x_e f_P(T/m_e)\right].\label{eq: P(r,t)}
\end{align}	

The last thing left to define in Eqs.~(\ref{eq:N-S 1}-\ref{eq: N-S 3}) is the $\dot{q}(r,t)$ term in the third equation, which corresponds to the rate of heating (for $\dot{q}> 0$) or cooling (for $\dot{q}<0$) per unit volume. This is given by
\begin{equation}
	\dot{q}=\frac{4 x_E \sigma_{\rm T} \rho_{\rm cmb}}{m_p m_e} (T_{\rm cmb}-T)\rho,
\end{equation}
where $\sigma_{\rm T}\simeq 6.65 \times 10^{-25}~\text{cm}^2$ is the Thomson scattering cross section, $\rho_{\rm cmb}\simeq (1.97\times 10^{-15}~\text{eV}^4)(1+z)^4$  is the CMB energy density, and $T_{\rm cmb}\simeq (2.34 \times 10^{-4}~\text{eV})(1+z)$ is the CMB temperature at redshift $z$. 

Using the system of equations in Eqs.~(\ref{eq:N-S 1}-\ref{eq: N-S 3}) completely defines the evolution of baryonic matter around the EDO. However, in practice, to handle these equations we still need to take a set of approximations that will make it possible to perform this calculation. These are the following:
\begin{enumerate}
	\item We will focus on obtaining a static approximation, for which we can neglect the time evolution of all the parameters defined so far. In this way, we will just calculate the solutions for the functions $\rho(r)$, $T(r)$, $P(r)$, $n_e(r)$, $x_e(r)$, $\mathcal{E}(r)$ and $v(r)$. 
	\item Taking the \textit{hydrostatic approximation}, we can take the baryon flow to vanish, such that $v(r)=0$. This is a valid approximation as long as the gravitational interaction is not too intense, placing limits on EDOs with a radius of the same order as their Schwarzschild radius.
	\item We will set $x_e(r)$ to be constant when solving for the Navier-Stokes equations, and will add the effects of the ionisation  of the baryon fluid as a correction to their solution. 
	\item Taking the ratio between the internal energy, $\mathcal{E}(r)$, and pressure, $P(r)$, we obtain the adiabatic index, $\gamma(r)$, 
	\begin{equation}\label{eq: E-P ratio}
		\frac{\mathcal{E}(r)}{P(r)}=\gamma(r)-1,
	\end{equation}
	which for $x_e\ll1$, gives $\gamma(r)\approx 5/3$, while for $x_e\approx1$ we have $\gamma(r)\approx5/3 $ if $T(r)\ll m_e$ and $\gamma(r)\approx13/9$ if $T(r)\gg m_e$. Therefore, we will approximate this function to be always constant except for $T(r)=m_e$, such that
 \begin{equation}\label{eq: gamma(r)}
     \gamma(r)=
     \begin{cases}
         \frac53  & \quad  T(r) \leq m_e \\
         \frac{13}{9} & \quad  T(r)>m_e.
     \end{cases}
 \end{equation}
  Doing so, allows us to relate $P(r)$ and the baryon density, $\rho(r)$ as follows
	\begin{equation}
		P(r)=K\rho(r)^\gamma,
	\end{equation}
	where $K$ is a constant of mass dimension $4-4\gamma$. Inserting this relation into Eq.~\eqref{eq: P(r,t)} leads to
	\begin{equation}
		T(r)=K m_p\frac{\rho(r)^{\gamma-1}}{1+x_e f_P},
	\end{equation}
	where in order for $T(r)$ to be continous at $T(r_{\rm rel})=m_e$ we will have different, but constant, values for $K,\gamma$ and $f_P$ in each side of that boundary.
\end{enumerate}

Considering this set of approximations, the Navier-Stokes equations simplify to
\begin{equation}\label{eq: Main DE}
	\frac{GM(r)}{r^2}+\gamma K \rho(r)^{\gamma-2} \frac{d \rho(r)}{dr}=0.
\end{equation}

\subsection{EDO Mass profiles}
The only thing left to define is the mass function, $M(r)$, which will vary depending on the formation mechanism for the EDO. In this work, we will consider the following mass profiles:
\begin{itemize}
    \item \textbf{Uniform spheres (of constant density):} Although this is a toy model, this mass profile describes quark nuggets, either from standard~\cite{Witten:1984rs} or axion~\cite{Zhitnitsky:2002qa} QCD. The mass profile is given by 
    \begin{equation}
    M(r)=M \begin{cases}
        \left(\frac{r}{R}\right)^3 &\qquad r\leq R\\
        1 &\qquad R<r.
    \end{cases}
    \label{eq:uniformsphere}
    \end{equation}
    \item \textbf{Boson stars:} Obtained by the gravitational collapse of a scalar field~\cite{Schunck:2003kk}. They do not have an analytical mass function, but they can be obtained by solving Schrodinger's equation for a scalar field under the influence of its gravitational potential~\cite{Bar:2018acw}. As the solution converges at infinity, we will define the radius where it encloses $90\%$ of its mass.
    \item \textbf{Navarro-Frenk-White (NFW) subhalos:} Empirically obtained density distribution that closely matches the behaviour of cold dark matter clustering~\cite{NFW}. The mass profile is given by 
    \begin{equation}
        M(r)=\int_0^{r}\d \hat{r}\,4\pi\hat{r}^2 \rho_{\rm NFW}(\hat{r}),
        \label{eq:nfw}
    \end{equation}
    where
    \begin{equation}
        \rho_{\rm NFW}(\hat{r})=\frac{\rho_0}{\frac{\hat{r}}{R_s}\left(1+\frac{\hat{r}}{R_s}\right)^2},
    \end{equation}
    is the NFW density profile, with $R_s$ and $\rho_0$ being parameters that define the total subhalo mass and $r_{90}$, the radius that encloses $90\%$ of its mass. {Because the mass of this profile technically diverges, we choose to cut it off at $100 R_s$.}
    \item \textbf{Ultra-compact mini-halos:} Although it is widely known that primordial density fluctuations seed the creation of primordial black hole, this fact depends on the gravitational pull of the over-density. Thus, for small enough over-densities that do not form a pbh, it is still possible to seed the growth of an ultra-compact mini-halo~\cite{Ricotti:2009bs}. The mass profile of these objects is expected to be~\cite{Bertshinger}
    \begin{equation}
    M(r)=M \begin{cases}
        \left(\frac{r}{R}\right)^{3/4} &\qquad r\leq R\\
        1 &\qquad R<r.
    \end{cases}
    \label{eq:ucmh}
    \end{equation}
\end{itemize}

\begin{figure}
    \centering
    \includegraphics[width=0.5\textwidth]{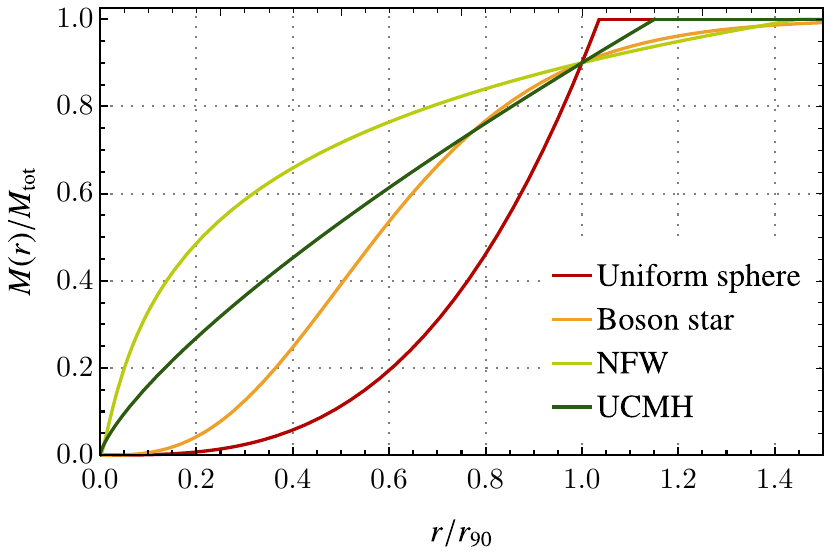}
    \caption{Mass profiles of the EDOs described in text.}
    \label{fig:mass_profiles}
\end{figure}

We show the different mass profiles in Figure~\ref{fig:mass_profiles}. With this, we are all set to solve Eq.~\eqref{eq: Main DE} and see the impact that EDOs can have on the ionisation  history of the universe. In the next subsection, we will proceed to obtain the density, temperature, ionisation  fraction and electron/proton number density of the accreted baryonic matter for a generic EDO.

\subsection{Accretion profiles of baryonic matter}
As introduced above, we solve our system of equations numerically, which, although it is less efficient than the analytical route from Ref.~\cite{Bai:2020jfm}, allows extending our results to any mass function. Moreover, in Appendix~\ref{App code} we present how this computation can be optimised so that solutions for a specific EDO with a given total mass and radius can be easily generalised to a different one.
 
Solving for the density profile of the accreted baryons around the extended dark matter objects will first require solving Eq.~\eqref{eq: Main DE} both inside and outside the EDO, matching both solutions at the boundary. To do so, we also need to specify the boundary conditions at infinity, defined as
\begin{align}
	\lim_{r\to\infty}\rho(r) \equiv \, &\rho_\infty & \lim_{r\to\infty}T(r) \equiv \, &T_\infty & \lim_{r\to\infty}x_e(r)\equiv \, &\bar{x}_e,
\end{align}
which will be defined by the homogeneous evolution of the Universe, as we will discuss in section~\ref{sec: ionisation }. Using Eqs.~\eqref{eq: P(r,t)} and \eqref{eq: gamma(r)}, and the fact that $T_\infty\ll m_e$, we further obtain
\begin{align}
	P_\infty=\frac{1}{m_p}T_\infty\rho_\infty\left[1+\bar{x}_e\right], \qquad\text{and} \qquad \gamma_\infty=5/3.
\end{align}
From this, one can find a specific form for $\rho(r)$, which is extended to the temperature and electron/proton number density using
\begin{align}
	T(r)=&K m_p \frac{\rho(r)^{\gamma-1}}{1+\bar{x}_e}\label{eq: T(r) vanilla}\\
	n_e(r)=&\frac{1}{m_p}\bar{x}_e \rho(r),
\end{align}
where we have defined $K=T_\infty/\rho_\infty^{\gamma_\infty-1}$ from the boundary condition at infinity.
However, this solution is incomplete until we include the thermodynamic corrections that lead to the cooling due to interactions with the CMB radiation, the ionisation  of the fluid, and the relativistic transition of electrons. We will now explain how to account for each of these corrections to our solutions.

\subsubsection{CMB cooling}
One of the first assumptions we made in the previous subsection was setting the the baryon fluid heating rate to zero (i.e., $\dot{q}=0$). Here we will allow for different values of this function, and check its impact on the temperature obtained through Eq.~\eqref{eq: T(r) vanilla}.

    Starting from the static version for Eqs.~(\ref{eq:N-S 1}) and (\ref{eq: N-S 3}),
    \begin{align}
         \frac{1}{r^2}(r^2\rho v )'&=0,\\
  \rho v (\mathcal{E}/\rho)' +P \frac{1}{r^2}&(r^2v)'=\dot{q},
    \end{align}
    and using Eqs.~\eqref{eq: P(r,t)} and \eqref{eq: E-P ratio}, we obtain
    \begin{equation}\label{eq: cooling}
        v \rho^{\gamma_\infty -1} \frac{\d }{\d r}\left(\frac{T}{\rho^{\gamma_\infty-1}}\right)=\frac{8 \bar{x}_e \sigma_{\rm T} \rho_{\rm cmb}}{3m_e(1+\bar{x}_e)} (T_{\rm cmb}-T),
    \end{equation}
    where we have also set the values at infinity for $\gamma(r)$ and $x_e(r)$. To analyze the different limits of this differential equation, let us first introduce two quantities that derive from the sound speed of the baryonic fluid at infinity,
    \begin{equation}
        c_\infty=\sqrt{\gamma_\infty P_\infty/\rho_\infty}.
    \end{equation}
    Defining the Bondi radius~\cite{Bondi:1952ni} as the characteristic distance where the EDO influences the baryon fluid, we obtain
    \begin{equation}\label{eq: Bondi radius}
    R_{\rm B}=GM/c_\infty^2,
    \end{equation}
    which is related to the escape velocity of the EDO. Similarly, the Bondi time can be defined as the time it takes a perturbation of the baryon fluid to cover the Bondi radius, leading to
    \begin{equation}
        t_{\rm B}=GM/c_\infty^3.
    \end{equation}
    With this, we can now introduce these characteristic scales in Eq.~\eqref{eq: cooling} to define the impact of cooling on the EDO formation. Multiplying $t_{\rm B}$ on both sides, we get
    \begin{equation}\label{eq: cooling bondi}
        \frac{v}{c_\infty} \rho^{\gamma_\infty -1} R_{\rm{B}}\frac{\d }{\d r}\left(\frac{T}{\rho^{\gamma_\infty-1}}\right)=\Theta(T_{\rm cmb}-T),
    \end{equation}
    where following Ref.~\cite{Bai:2020jfm} we define the dimensionless cooling factor as
    \begin{equation}
        \Theta=\frac{8 \bar{x}_e \sigma_{\rm T} \rho_{\rm cmb}}{3m_e(1+\bar{x}_e)} t_{\rm B}.
    \end{equation}
    For the region of interest, outside of the EDO, we expect both {$R_{\rm B}\d/{\d r} (T/\rho^{\gamma_\infty -1})/(T/\rho^{\gamma_\infty -1})$} and $v/c_\infty$ to be $\mathcal{O}(1)$. Thus, we can see two extreme cases for this differential equation: for $\Theta\ll1$, the right-hand side vanishes, recovering the simplified temperature profile described in Eq.~\eqref{eq: T(r) vanilla}. On the contrary, when $\Theta\gg1$, the only fixed point for Eq.~\eqref{eq: cooling bondi} is at $T(r)=T_{\rm cmb}$, cooling down the EDO due to the interactions with the CMB. By continuity, there must be a threshold for which the cooling effect due to the CMB only partially affects the temperature of the accreted matter.
    In Ref.~\cite{Ali-Haimoud:2016mbv}, the authors obtain an estimate for this boundary of $r_{\rm cool}=\Theta^{-2/3}R_{\rm B}$ using the analytic solution for PBHs. However, we can arrive at the same conclusion by studying Eq.~\eqref{eq: cooling bondi} itself. Focusing on the cooling factor, we can see that the aforementioned two extreme cases can be also expressed in terms of a cooling time (defined as the inverse of the cooling rate), given by
    \begin{equation}
        t_{\rm cool}=\frac{3m_e(1+\bar{x}_e)}{8 \bar{x}_e \sigma_{\rm T} \rho_{\rm cmb}}.
    \end{equation} 
    For $t_{\rm cool}\gg t_{\rm B}$, accreted matter has enough time to escape the EDO before the CMB cools it down, leading to a negligible effect on its temperature profile. However, when $t_{\rm cool}\ll t_{\rm B}$, the accreted matter cools down before it crosses the Bondi radius, which {effectively} shrinks the radius at which the fluid can escape. This new effective radius, $r_{\rm cool}$, can be obtained following the same arguments as for the Bondi radius, such that
    \begin{equation}
        r_{\rm cool} {\equiv} \frac{GM}{c_{\rm cool}^2}=\frac{GM t_{\rm cool}^2}{r_{\rm cool}^2}= \frac{GM}{c_\infty^2}c_\infty^2\frac{t_{\rm cool}^2}{r_{\rm cool}^2}=\frac{R_{\rm B}^3}{r_{\rm cool}^2 \Theta^2},
        \label{eq:Rcool}
    \end{equation}
    where $c_{\rm cool}=r_{\rm cool}/t_{\rm cool}$ is the adiabatic sound speed of the cooled fluid. Eq.~\eqref{eq:Rcool} can be solved to give $r_{\rm cool}=\Theta^{-2/3}R_{\rm B}$. Therefore, when adding the cooling effect, the matter temperature will be given by
    \begin{equation}
        T(r)=\begin{cases}
            T_{\rm cmb}\qquad &r\geq r_{\rm cool}\\
            T_{\rm cmb}\left(\frac{\rho(r)}{\rho(r_{\rm cool})}\right)^{\gamma-1}\qquad &r<r_{\rm cool},
        \end{cases}
    \end{equation}
    where $\rho(r)$ is obtained by solving Eq.~\eqref{eq: Main DE}. In our analysis, we will incorporate this effect by shifting $R_{\rm B}\to r_{\rm cool}$ for values of $\Theta>1$, as this is when the effect becomes relevant. 
    
    \subsubsection{Ionisation of matter}
    Independently of the ionisation  fraction at infinity, $\bar{x}_e$, the rising energy of the accreted matter inside the EDO can provide enough energy for the electrons in the hydrogen atoms to escape via scattering processes such as $H+H\to H+e^-+p+\gamma$, what is known as \textit{collisional ionisation}. The minimum temperature at which this effect comes into play depends on the properties of the fluid at infinity; however, following Ref.~\cite{Bai:2020jfm}, we will take it to be $T_{\rm ion}\simeq 1.5\times 10^4~\text{K}\approx1.3~\text{eV}$~\cite{Nobili:1991tx}. 

    Therefore, 
    the temperature of the fluid will evolve following
    Eq.~\eqref{eq: T(r) vanilla} until it reaches the first threshold, $r_{\rm start}$, given by $T(r_{\rm start})=T_{\rm ion}=1.3~\text{eV}$. At that point, the accreted matter collisionally ionizes until $r_{\rm end}$, defined by $x_{e}(r_{\rm end})=1$. Inside this region of space, the temperature of the fluid can be taken to remain constant as this process only redistributes the internal energy of the gas, while the evolution of the ionisation  fraction $x_e(r)$ can be calculated using the first law of thermodynamics, leading to 
    \begin{equation}
        \frac{\d}{\d r}\left(\frac{3}{2}[1+x_e(r)]T(r)-[1-x_e(r)]E_{\rm I}\right)=-[1+x_e(r)]T(r)\rho(r)\frac{\d \left(1/\rho(r)\right)}{\d r},
    \end{equation}
    where $E_{\rm I}\simeq13.6~\text{eV}$ is the binding energy of the neutral hydrogen. Solving this equation gives
    \begin{equation}
        x_e(r)=(1+\bar{x}_e)\left(\frac{\rho(r)}{\rho(r_{\rm start})}\right)^{\left(\frac{3}{2}+\frac{E_{\rm I}}{T_{\rm ion}}\right)^{-1}}-1,
    \end{equation}
    where the $\rho(r_{\rm start})$ denominator ensures the matching with boundary condition {$x_e(r_{\rm start})=\bar{x}_e$}. Once $x_e(r)$ reaches $x_e(r_{\rm end})=1$, the temperature of the accreted matter unfreezes and evolves similarly to Eq.~\eqref{eq: T(r) vanilla}, ensuring continuity. Therefore, we can account for collisional ionisation  as follows:
    \begin{align}
        T(r)=&\begin{cases}
        T_{\rm cmb}\qquad &r\geq r_{\rm cool}\\
        T_{\rm cmb}\left(\frac{\rho(r)}{\rho(r_{\rm cool})}\right)^{\gamma-1}\qquad &r_{\rm cool}\geq r>r_{\rm start}\\
            T_{\rm ion}\qquad & r_{\rm start}\geq r\geq r_{\rm end}\\
            T_{\rm ion}\left(\frac{\rho(r)}{\rho(r_{\rm end})}\right)^{\gamma-1}\qquad &r_{\rm end}<r\\
        \end{cases}\\\nonumber\\
        x_e(r)=&\begin{cases}
            \bar{x}_e\qquad &r>r_{\rm start}\\
            (1+\bar{x}_e)\left(\frac{\rho(r)}{\rho(r_{\rm start})}\right)^{\left(\frac{3}{2}+\frac{E_{\rm I}}{T_{\rm ion}}\right)^{-1}}-1\qquad &r_{\rm start}\geq r\geq r_{\rm end}\\
            1\qquad &r_{\rm end}>r.
        \end{cases}
    \end{align}
    We expect the photons produced through collisional ionisation also to ionize nearby neutral atoms when their energy is higher than the binding energy, $E_{\rm I}$. This is called {\textit{photoionisation }~\cite{2006agna.book.....O}}, and is a second source of ionisation  that can take place inside EDOs. Realistically, both processes would occur simultaneously, but it is common practice to calculate both cases separately, obtaining higher and lower bounds for the system. Since the most conservative results are given by collisional ionisation, we will just focus on that case in our analysis. 
    
    \subsubsection{Relativistic electrons}
    Depending on the specific EDO properties, it is possible to heat the accreting matter enough so that electrons become relativistic, for which we will take $T(r)\geq2 m_e/3$. As introduced in section~\ref{sec: Thermo}, the relativistic nature of electrons has an impact on the pressure and the internal energy density of the baryon fluid, controlled by the functions $f_{\mathcal{E}}(T/m_e)$ and $f_{P}(T/m_e)$. However, in terms of the differential equation for the density profile in Eq.~\eqref{eq: Main DE}, this difference arises through $\gamma$, which, as defined in Eq.~\eqref{eq: gamma(r)}, evaluates to $5/3$ for non-relativistic temperatures, and $13/9$ otherwise. 

\begin{figure}
        \centering
        \includegraphics[width=0.9 \textwidth]{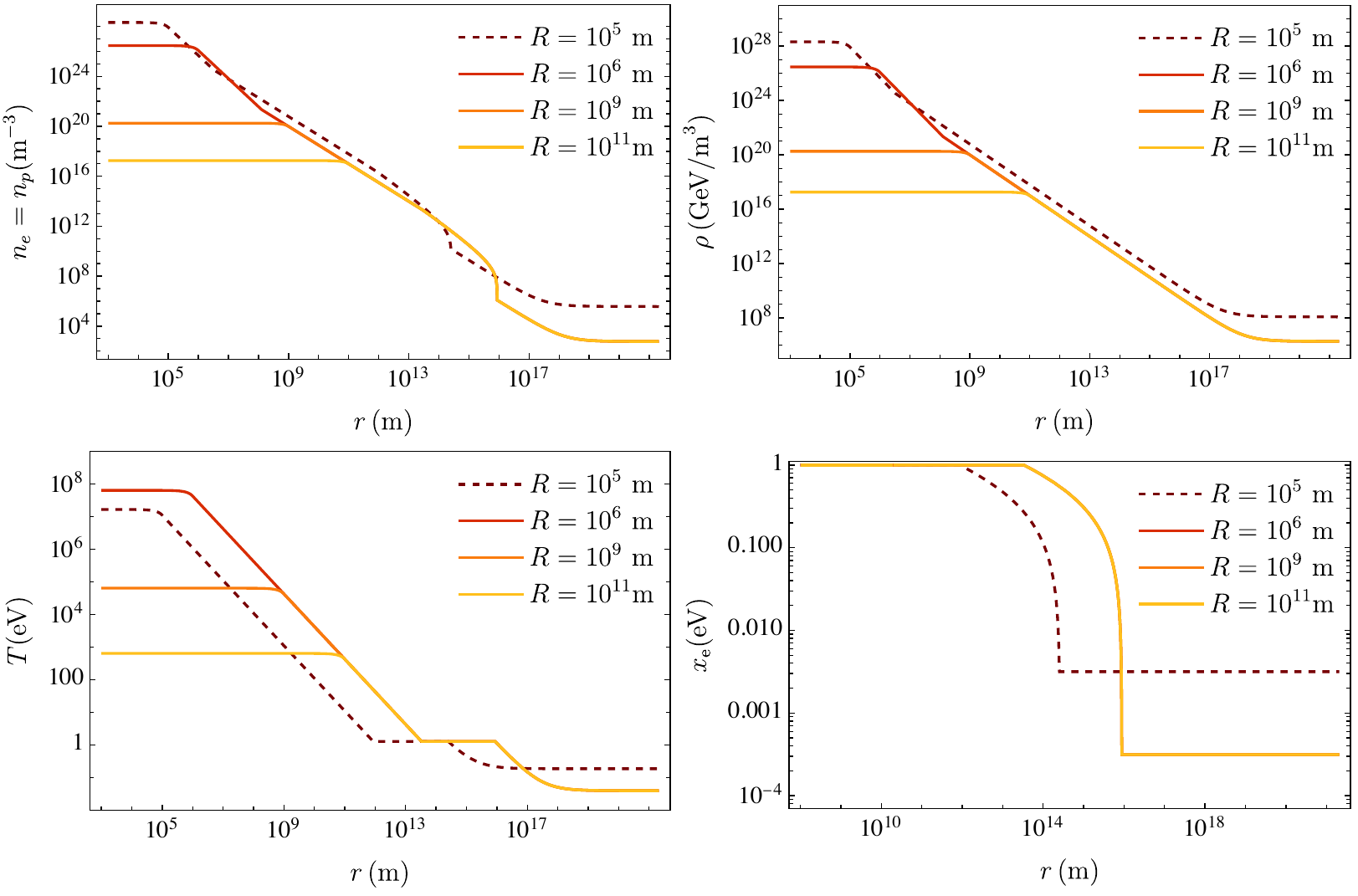}
        \caption{Solutions for the electron/proton number density (top-left), energy density (top-right), temperature (bottom-left) and ionisation  fraction (bottom-right) for a uniform EDO of mass ${M=10^5~M_\odot}$ but different radii. The dashed solution represents an EDO at $z=800$, while plain corresponds to $z=200$, as given in Eq.~\eqref{eq: boundary conditions inf}. We can see the effect that collisional ionisation  has on temperature, and how the CMB cooling delays the heating of the accreted matter for the $z=800$ solution. 
        }
        \label{fig:Uniform-Profiles}
    \end{figure}

\begin{figure}[h]
        \centering
        \includegraphics[width=0.9 \textwidth]{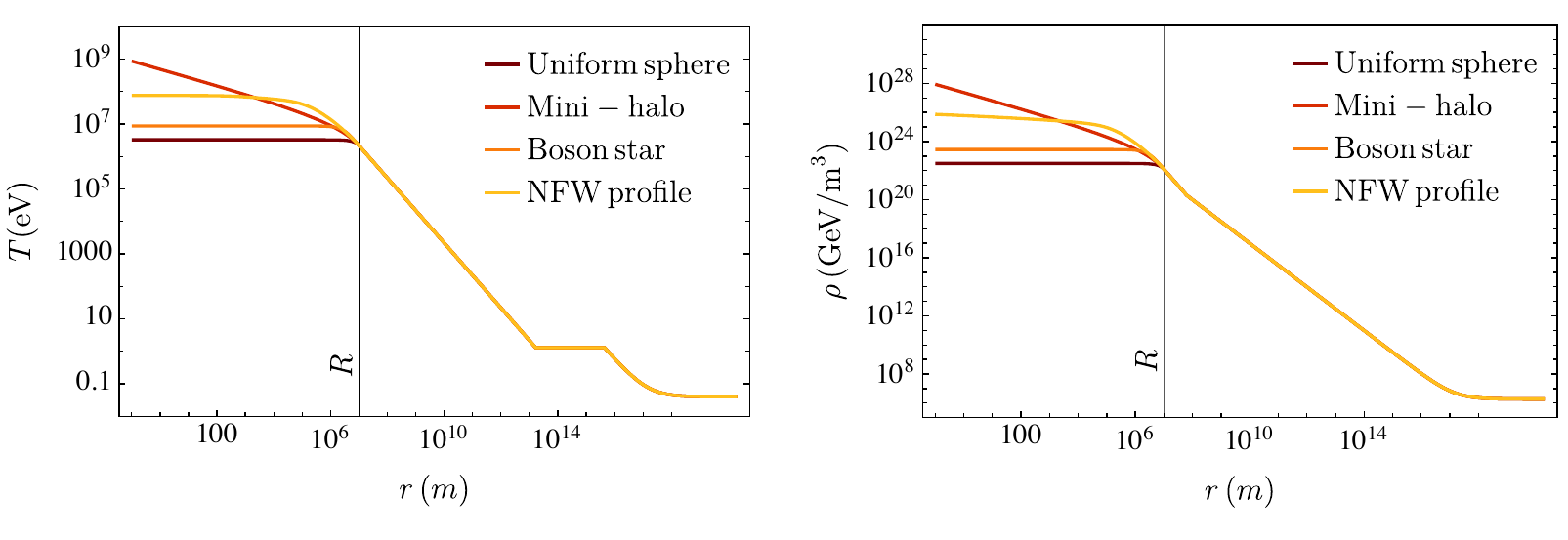}
        \caption{Temperature (left) and energy density (right) of accreted matter into EDOs with same total mass ($M=10^5M_\odot$) but different mass functions. As expected, all profiles agree up to the EDO radius, $R=10^7\,{\rm m}$, after which each function leads to a different behaviour for the accreted matter.}
        \label{fig:profiles_mass_fns}
    \end{figure}

   Moreover, as we can see in Eq.~\eqref{eq: T(r) vanilla}, the temperature relation to density also depends on the adiabatic factor $\gamma$. However, this dependence perfectly cancels the one in the differential equation from Eq.~\eqref{eq: Main DE}, implying that the evolution of the temperature is not affected by relativistic effects. Therefore, the only thing that we need to address is the energy density, for which we can take two different routes: use Eq.~\eqref{eq: Main DE} to solve the evolution of $\rho(r)$ for the relativistic region numerically, or use the already calculated value for $T(r)$, and rescale it to $\rho(r)$ using the correct value for $\gamma=13/9$ inside the relativistic radius, $r_{\rm rel}$. Taking the second route, the relativistic effects are incorporated as follows:
\begin{align}
        T(r)=&\begin{cases}
        T_{\rm cmb}\qquad &r\geq r_{\rm cool}\\
            T_{\rm cmb}\left(\frac{\rho_{\rm sol}(r)}{\rho_{\rm sol}(r_{\rm cool})}\right)^{\gamma-1}\qquad &r_{\rm cool}\geq r>r_{\rm start}\\
            T_{\rm ion}\qquad & r_{\rm start}\geq r\geq r_{\rm end}\\
            T_{\rm ion}\left(\frac{\rho_{\rm sol}(r)}{\rho_{\rm sol}(r_{\rm end})}\right)^{\gamma-1}\qquad &r_{\rm end}<r\\
        \end{cases}\label{eq:T final}\\\nonumber\\
        x_e(r)=&\begin{cases}
            \bar{x}_e\qquad &r>r_{\rm start}\\
            (1+\bar{x}_e)\left(\frac{\rho_{\rm sol}(r)}{\rho_{\rm sol}(r_{\rm start})}\right)^{\left(\frac{3}{2}+\frac{E_{\rm I}}{T_{\rm ion}}\right)^{-1}}-1\qquad &r_{\rm start}\geq r\geq r_{\rm end}\\
            1\qquad &r_{\rm end}>r
        \end{cases}\\\nonumber\\
        \rho(r)=&\begin{cases}
            \rho_{\rm sol}(r)&r>r_{\rm rel}\\
            \rho_{\rm sol}(r_{\rm rel}) \left(\frac{\rho_{\rm sol}(r)}{\rho_{\rm sol}(r_{\rm rel})}\right)^{3/2}&r_{\rm rel}\geq r,
        \end{cases}
    \end{align}
    where we have called $\rho_{\rm sol}(r)$ to the solution obtained by solving Eq.~\eqref{eq: Main DE} for $\gamma=5/3$. Having this, the number of electrons is given by computing Eq.~\eqref{eq: ne np} with the obtained functions, such that
     \begin{equation}\label{eq: ne final}
        n_e(r)=\frac{1}{m_p}x_e(r)\rho(r).
    \end{equation}
    In Figure~\ref{fig:Uniform-Profiles} we show some examples of the density, temperature, ionisation  fraction, and number of electrons for constant-density EDOs with various radii and masses. Additionally, in Figure~\ref{fig:profiles_mass_fns}, we show a comparison of the effect that different mass functions have on the accreted matter.
    
Taking the series of approximations and correcting for thermodynamic effects we have thus obtained the static behaviour of a baryonic fluid around dark matter compact objects. In particular, we expect the heating and ionisation of the accreted matter to have an impact on the thermal history of the universe. In the next section, we will study EDOs in this context, placing constraints on their formation.

\section{Ionisation  of the cosmic microwave background}\label{sec: ionisation }
Before addressing the effect that EDOs have on the ionisation history of the universe, we first need to define the boundary conditions at infinity for the EDOs. Here we will use the Peebles Case B~\cite{PeeblesCaseB,Poulin:2015pna} model of recombination, also called the \textit{three-level atom} model, to obtain the background evolution of baryon temperature ($T_{\rm M}$) and ionisation  fraction ($x_e$). This is described by the set of equations\footnote{We recommend Ref.~\cite{Baumann:2022mni} for a pedagogical introduction to these equations.}
 \begin{align}\label{eq:Peebles vanilla T}
     \frac{\d T_{\rm M}}{\d z}&=\frac{1}{(1+z)}\left[2T_{\rm M} +\frac{8\pi^2 \sigma_{\rm T} {T_{\rm cmb}}^4}{45H(z)m_e}\frac{x_e}{1+x_e}(T_{\rm M}-T_{\rm cmb})\right],\\
     \frac{\d x_e}{\d z}&=C_r(z)\frac{\alpha_{\rm B}(T_{\rm M})}{H(z)(1+z)}\left[n x_e^2 +\left(\frac{m_e T_{\rm cmb}}{2\pi}\right)^{3/2}e^{-\frac{E_{\rm I}}{T_{\rm cmb}}}(1-x_e)\right].\label{eq:Peebles vanilla xe}
 \end{align}
 All the necessary definitions are described in what follows. $H(z)$ is the Hubble parameter, where we will take Planck's value for $H_0=67.44$ km/Mpc/s. $\sigma_{\rm T}\simeq 6.65\times10^{-25}\,{\rm cm}^2$ is the Thomson's constant, $E_{\rm I}=13.2$ eV is the binding energy for the ground state of the hydrogen atom, $T_{\rm cmb}\simeq2.35\times 10^{-4}\,(1+z)\,{\rm eV}$ is the already mentioned CMB radiation temperature and $n(z)=\rho(z)/m_{\rm p}$ is the baryon density number. Finally, $C_r(z)$ is the effective branching ratio (or the so-called Peebles factor), and it describes the probability that excited electrons in a hydrogen atom will reach the ground state. This factor is given by
 \begin{equation}
     C_r(z)=\frac{1+K_{H}\Lambda_{H}n(1-x_e)}{1+K_H(\Lambda_H+\beta_\alpha(T_{\rm M}))n(1-x_e)}
 \end{equation}
 where $K_H={\lambda_{\rm Ly}}^3(8\pi H(z))$ with $\lambda_{\rm Ly}=121.5$ nm being wavelength of a Lyman-$\alpha$ photon, $\Lambda_H=8.227~\text{s}^{-1}$ is is the two-photon decay rate of the hydrogen 2S state, and
 \begin{align}
     \beta_\alpha(T_{\rm M})&=\alpha_{\rm B}(T_{\rm M})\left(\frac{m_e T_{\rm cmb}}{2\pi}\right)^{3/2}e^{-E_{\rm I}/4T_{\rm cmb}},\\
 \alpha_{\rm B}(T_{\rm M})&=9.8 \frac{\alpha^2}{m_e^2}\left(\frac{E_{\rm I}}{T_{\rm M}}\right)\log\left(\frac{E_{\rm I}}{T_{\rm M}}\right)
 \end{align}
 are characteristic parameters for the ionisation  and recombination rates, respectively, with $\alpha=1/137$ being the fine-structure constant. Considering higher energy levels of hydrogen and other atom species in the universe requires an extension of these equations that will add sub-percentage corrections to these equations~\cite{2007MNRAS.374.1310C,2011PhRvD..83d3513A,2013ascl.soft04017C,2011MNRAS.412..748C}, but these are beyond the error estimates of this work.

 Therefore, we will use the solution for this system of equations as our boundary conditions at infinity of the EDOs, such that
 \begin{align}\label{eq: boundary conditions inf}
     \rho_\infty&\equiv m_p n(z), & T_\infty\equiv & ~T_{\rm M}(z), & \bar{x}_e\equiv x_e(z).
 \end{align}
 However, if an EDO is placed in this background, we expect it to backreact onto the baryon fluid, modifying the recombination history as given by Eqs.~(\ref{eq:Peebles vanilla T}-\ref{eq:Peebles vanilla xe}). For this, we will calculate the luminosity and the energy deposition rate of these objects into the background, allowing us to obtain the modifications to the ionisation fraction of the universe.
 
 \subsection{Luminosity and energy deposition rate}
 For a spherically symmetric extended dark matter object such as the ones described in section~\ref{sec:accretion}, in-falling electrons and protons may scatter in such a way that can lead to bremsstrahlung radiation via processes such as $ee\to ee\gamma$ or $ep\to ep\gamma$. The emissivity of this radiation, defined as the emission power per volume per frequency per steradian, into photons of frequency $\nu$ at a distance $r$ of the EDO is given by
 \begin{equation}
     j_\nu(r)=\frac{8}{3}\left(\frac{2\pi m_e}{3T(r)}\right)^{1/2}\frac{\alpha^3}{m_e^2}g_{ff}(\nu,T(r))e^{-2\pi \nu/T(r)}n_e(r)n_p(r),
 \end{equation}
 where $g_{ff}(\nu,T(r))$ is the free-free Gaunt factor that adds in the quantum corrections, and $T(r)$ and $n_e(r)=n_p(r)$ are given by Eqs.\eqref{eq:T final} and \eqref{eq: ne final}, respectively.

Integrating the frequency in the emissivity we obtain the radiation power density, defined as the radiation per power per volume, which can be expressed as~\cite{1983MNRAS.204.1269S}
\begin{equation}\label{eq: L(r)}
    \mathcal{L}(r)=n_e(r)n_p(r)\alpha \sigma_{\rm T} T(r) \mathcal{J}(T(r)/m_e),
\end{equation}
where $\mathcal{J}(T(r)/m_e)$ is a dimensionless function encoding the information from the free-free Gaunt function. Considering both the formulas for $e-p$~\cite{e-pPlasma} and $e-e$~\cite{e-ePlasma} free-free emissivities, Ref.~\cite{Ali-Haimoud:2016mbv} provides a percent accuracy level approximation given by
\begin{equation}
    \mathcal{J}(X)\approx
    \begin{cases}
        \frac{4\sqrt{2}}{\pi\sqrt{\pi}}X^{-1/2}(1+5.5X^{1.25}), \qquad &X<1,\\
        \frac{27}{2\pi}\left[\log\right(2Xe^{-\gamma_{\rm E}}+0.08\left)+\frac{4}{3}\right],\qquad &X>1,
    \end{cases}
\end{equation}
where $\gamma_{\rm E}=0.557$ is the Euler Gamma constant. Finally, by integrating over volume the radiation power density we obtain the total luminosity. However, using Eq.~\eqref{eq: L(r)} to calculate the luminosity will give us an overestimate, since we are only interested in the luminosity difference due to the addition of the EDO. Therefore, we will use the following definition of the EDO accretion luminosity,
\begin{equation}
    L=\int_0^\infty \d r \, 4 \pi r^2\left[\mathcal{L}(r)-\mathcal{L}(\infty)\right],
\end{equation}
which vanishes in the absence of an extended dark matter object. Notice that our static solution for the accreted matter into an EDO is consistent only up to $r=R_B$, placing an upper cut-off to this integral. Finally, the power density at redshift $z$ is given by
\begin{equation}
    P(z)=L(z) n_{\rm EDO}(z),
\end{equation}
where the redshift dependence comes from the boundary conditions at infinity as defined in Eq.~\eqref{eq: boundary conditions inf}, and $n_{\rm EDO}(z)$ is the number density of extended dark objects at a given redshift, defined as $n_{\rm EDO}(z)=f_{\rm DM}\,\rho_{\rm DM}(z)/M$, where $\rho_{\rm DM}(z)$ is the density of dark matter and $f_{\rm DM}$ the fraction of dark matter covered by EDOs of same total mass $M$. 
\subsection{Depositing EDOs in our universe: relative distances and velocities}
So far, the solutions for extended dark matter objects we studied were simplified so that they were isolated and in the same frame of reference as the thermal bath. In this subsection, we will address these approximations, getting a more realistic solution for their effects on the CMB.

To make sure that the isolation approximation is valid, we need to compare the Bondi radius for the EDOs, defined in Eq.~\eqref{eq: Bondi radius}, with their average separation as a dark matter candidate, $d_{\rm s}$. Using the dark matter density, we find
\begin{equation}
    d_s(z)\sim \left(\frac{M }{\rho_{\rm DM}(z)f_{\rm DM}}\right)^{1/3}.
\end{equation}

Therefore, the condition $d_s(z)\gg R_{\rm B}$ translates into
\begin{equation}\label{eq single EDO assumption}
M\ll \sqrt{\frac{(\gamma_\infty T_\infty (z))^3}{(2Gm_p)^3 \rho_{\rm DM}(z) f_{\rm DM}}}\overset{z\to50}{\approx} \frac{3 \times10^4M_{\odot}}{\sqrt{f_{\rm DM}}},
\end{equation}
which places an upper limit in our analysis, depending on the dark matter fraction corresponding to EDOs. 

Taking into account relative velocities is more complicated. As one might expect, an EDO with a high relative speed compared to the speed of sound of baryon fluid, $v_{\rm rel}> c_s$, will be more evasive, leading to a smaller total accretion of matter and therefore emitted radiation. To deal with this scenario we will take the approach from Refs.\cite{Ali-Haimoud:2016mbv,Bondi:1952ni,1944MNRAS.104..273B}, where these effects are considered by shifting the speed of sound at infinity by $c_\infty\to\sqrt{c_\infty^2+v_{\rm rel}^2}$, decreasing the Bondi radius for the extended dark matter object. We thus expect $v_{\rm rel}$ to be a stochastic variable, and take it to be controlled by a three-dimensional Gaussian linear distribution with a standard deviation $\langle v_s^2\rangle^{1/2}$. \vtwo{For large relative speeds, it can be argued that the spherical accretion approximation should break down. Instead, disk-like accretion of baryonic matter should dominate above a certain threshold for $v_{\rm rel}$~\cite{Poulin:2017bwe, Serpico:2020ehh}. Including this effect would lead to tighter bounds on EDOs. In this work, we remain conservative and focus on this simplified system.}\footnote{We do not consider the effect of outflows. See~\cite{Bosch-Ramon:2020pcz} for a discussion about the appropriateness of accretion approximations in light of potential outflows, which may weaken the constraints, particularly for supersonic relative velocities.} Following Ref.~\cite{Ali-Haimoud:2016mbv}, the redshift dependence of the standard deviation is given by~\cite{Dvorkin:2013cea}
\begin{equation}
    \langle v_s^2\rangle^{1/2}=\min[1,z/10^3]\times 30 \text{km/s}.
\end{equation}
For a realistic estimation of the energy deposition rate of EDOs into the CMB radiation, we will have to calculate the weighted power density
\begin{equation}
    \langle P(z)\rangle=\langle L(z)\rangle n_{\rm EDO}(z),
\end{equation}
where the brackets include the effect of the relative velocities when applied on a function $\mathcal{O}$ as follows
\begin{equation}\label{eq: gaussian vrel}
    \langle\mathcal{O}\rangle=\frac{4\pi}{(2\pi\langle v_s^2\rangle/3)^{3/2}}\int_0^\infty \d v_{\rm rel}\, v_{\rm rel}^2 e^{-\frac{v_{\rm rel}^2}{2\langle v_s^2\rangle/3}}\mathcal{O}|_{c_\infty\to\sqrt{c_\infty^2+v_{\rm rel}^2}}.
\end{equation}
\begin{figure}
    \centering
    \includegraphics[width=\textwidth]{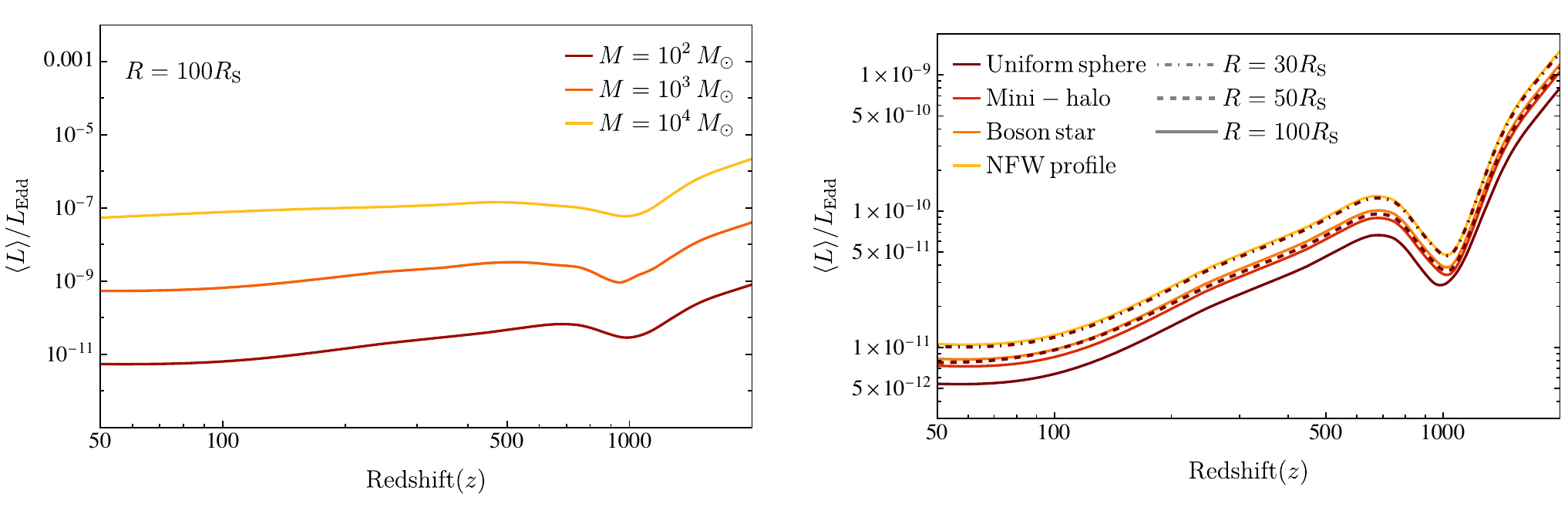}
    \caption{Relative-velocity averaged luminosity of a uniform EDO for different masses (left), normalized over the Eddington luminosity, given by $L_{\rm{Edd}}=4\pi G M m_p/\sigma_{\rm T}$, defined as the luminosity at which radiation pressure due Thomson scattering balances the accretion of hydrogen due to gravity. On the {right}, we plot the total averaged luminosity for the different EDO profiles with the same total mass $M=10^2M_\odot$, also normalized over the Eddington luminosity. Although the accretion profiles from Figure~\ref{fig:profiles_mass_fns} differed greatly inside the EDO radius, we can see that changing the mass function is {to very good approximation} equivalent to shifting the uniform sphere solution, as represented by the dashed and dot-dashed lines in the left-hand side plot.}
    \label{fig: Luminosity}
\end{figure}
Numerical calculations of this expression for the total luminosity {are computationally expensive. In}
Appendix~\ref{App code} we show how this process can be made more efficient by conveniently rescaling the radial components of the accreted matter profiles.

In Figure~\ref{fig: Luminosity} we show the relative velocity integrated luminosity for EDOs with different radii and mass functions. As we can see, the total luminosity from any mass function can be obtained {to very good approximation} by just shifting the uniform sphere solution. With this, we are ready to quantify how the accretion of matter into extended dark matter objects changes the ionisation  fraction of the background evolution.
\subsection{Effect on CMB anisotropies}
After recombination at redshift $z\sim1100$, it is assumed that most electrons and protons are combined into hydrogen, leading to an ionisation  fraction of $x_e\sim10^{-4}-10^{-3}$. This background neutrality allows the CMB radiation to move freely through space until today, except for the ignition of stars at $z\sim \mathcal{O}(10)$. Therefore, strong constraints 
 {can be placed using}
the state of the universe through the so-called \textit{dark ages}, as any deviation from neutrality would be a possible source for the absorption of photons, decreasing the visibility of CMB anisotropies. 
This is where the existence of EDOs comes into play, as their emitted radiation does not just overlay the CMB, but also heats the matter around it, ionizing the neutral hydrogen and locally causing $x_e\to1$. To place constraints, we need to find how this local change in the ionisation fraction affects the global evolution of the universe as studied in the Peebles case B recombination case in Eqs.~\eqref{eq:Peebles vanilla T} and \eqref{eq:Peebles vanilla xe}. As derived in Ref.~\cite{Ali-Haimoud:2016mbv}, the fraction of the power density that is deposited into the background, called the energy deposition rate, $\dot{\rho}_{\rm dep}(t)$, is given by\footnote{\vtwo{This analytical approximation uses the fact that, for the considered range of energies, the main process for photon cooling is via inelastic Compton scattering of electrons. Extending this calculation to allow for all interaction channels for photon cooling would add small corrections to our result, increasing the energy deposition rate and, thus, the obtained constraints. We also assume that all these interactions at a certain redshift occur on the spot.  Relaxing this approximation is not trivial and requires taking a numerical approach~\cite{Slatyer:2009yq,Slatyer:2015kla}.}}

 \begin{equation}
 a^{-7}\frac{\d}{\d t}\left(a^7\dot{\rho}_{\rm dep}\right)\approx 0.1\, n \,\sigma_{\rm T} (\langle P\rangle-\dot{\rho}_{\rm dep}),
 \end{equation}
 where the $0.1$ prefactor comes from the efficiency of the Compton scattering. We see that for $0.1\, n \,\sigma_{\rm T}a^7>1$ all the energy produced by the EDO gets deposited in the neutral background, until $z\approx200$, after which the energy injection decays as $a^{-7}$. 
 \begin{figure}
     \centering
     \includegraphics[width=\textwidth]{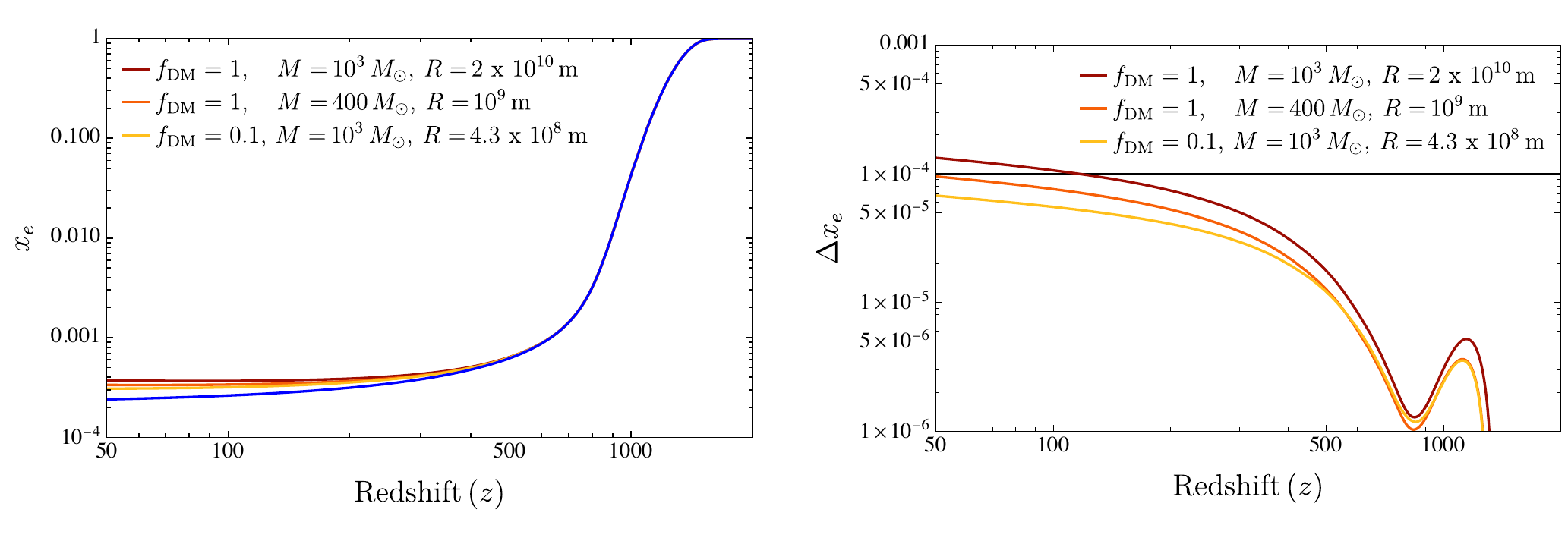}
     \caption{
     Modification of the ionisation  fraction due to the presence of uniform-sphere EDOs. On the left, the ionisation  fraction is presented against the $\Lambda$CDM evolution (blue) obtained solving Eqs.~(\ref{eq:Peebles vanilla T}-\ref{eq:Peebles vanilla xe}), while on the right we show the net modification for each set of parameters.
     }
     \label{fig: xes}
 \end{figure}
 
 Once we obtain the total energy deposited by EDOs at any redshift into the background, we are ready to see how they will impact the average temperature and ionisation  fraction of the universe. For this, we just need to modify the Peeble's equations from Eqs.~\eqref{eq:Peebles vanilla T} and \eqref{eq:Peebles vanilla xe} as follows
 \begin{align}\label{eq:Peebles EDO T}
     \frac{\d T_{\rm M}}{\d z}&=\frac{1}{(1+z)}\left[2T_{\rm M} +\frac{8\pi^2 \sigma_{\rm T} {T_{\rm cmb}}^4}{45H(z)m_e}\frac{x_e}{1+x_e}(T_{\rm M}-T_{\rm cmb})\right]-\frac{2}{3\,n}\frac{1+2x_e}{3H(z)(1+z)}\dot{\rho}_{\rm dep},\\
     \frac{\d x_e}{\d z}&=C_r(z)\frac{\alpha_{\rm B}(T_{\rm M})}{H(z)(1+z)}\left[n x_e^2 +\left(\frac{m_e T_{\rm M}}{2\pi}\right)^{3/2}e^{-\frac{E_{\rm I}}{T_{\rm M}}}(1-x_e)\right]-\frac{1-x_e}{3 H(z)(1+z)}\frac{\dot{\rho}_{\rm dep}}{E_{\rm I}\,n}.\label{eq:Peebles EDO xe}
 \end{align}

 In Figure~\ref{fig: xes}, we show how the ionisation fraction is modified by the addition of EDOs with different masses, radii, and dark matter fractions. To constrain these objects against their impact on the CMB anisotropies, we would have to implement the modified $x_e(z)$ in a Boltzman code and solve for the CMB power spectra, as it was done for PBHs in Refs.~\cite{Ali-Haimoud:2016mbv,Aloni:2016kuh}. 
 {However, following Ref.~\cite{Bai:2020jfm}, we note that we may also compare our EDO results to the existing results from PBHs without explicitly repeating this analysis.} \vtwo{This is because all accreting massive objects lead to similar imprints on the CMB independently of their shape or nature. Therefore, we can reuse previous constraints if we know how they modify the universe's ionisation fraction history. }

\vtwo{For this, using PBHs constraints from Ref.~\cite{Ali-Haimoud:2017rtz}, we can compare how different modifications to the ionisation fraction (Figure 12 in Ref.~\cite{Ali-Haimoud:2017rtz}) relate to the obtained constraints (Figure 14 in Ref.~\cite{Ali-Haimoud:2017rtz}). Moreover, as these modifications are largest for small redshift, as in Figure~\ref{fig: xes}, we focus on the modifications at $z=50$, which gives $\Delta x_e(z=50)<10^{-4}$, equivalent to having $\Delta x_e/x_e<\mathcal{O}(1)$ at all times.}\footnote{It is reasonable to expect that with better observations this threshold will be constrained to smaller values for {$\Delta x_e(z=50)$}. Since the corrections to the recombination equations from Eqs.~(\ref{eq:Peebles EDO T}-\ref{eq:Peebles EDO xe}) are subdominant, we can generally predict the impact that a change in $\Delta x_e(z=50)$ would have on the EDO constrains by just conveniently shifting $f_{\rm DM}$, as they are related by {$\delta \Delta x_e/\Delta x_e=\delta f_{\rm DM}/f_{\rm DM}$.}} 

\begin{figure}
    \centering
    \includegraphics[width=0.458\textwidth]{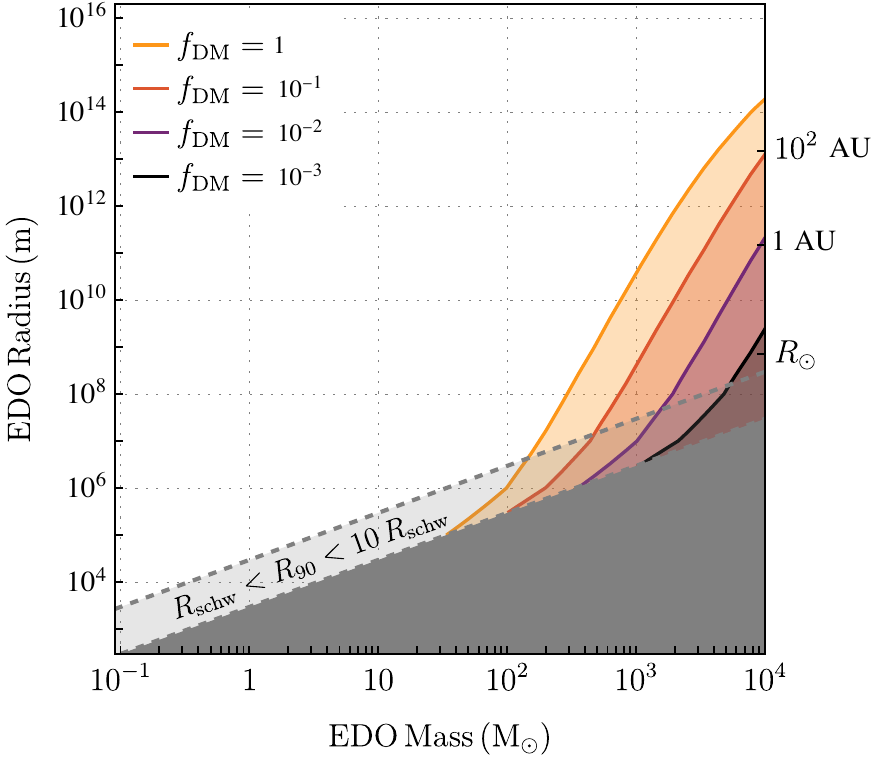}
    \includegraphics[width=0.425 \textwidth]{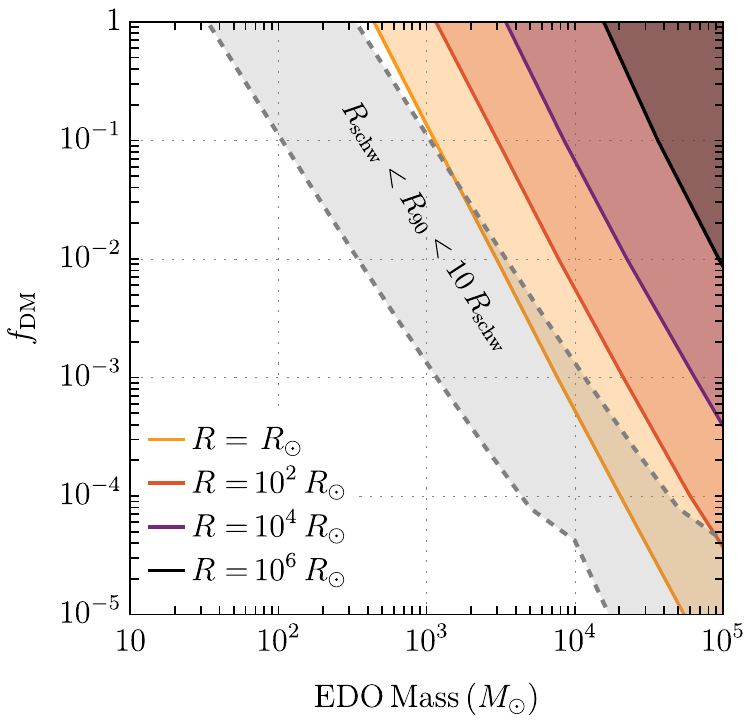}
    \caption{Constraints on uniform density EDOs for different dark matter fractions as a function of mass vs radius (right) and for different radii as a function of mass vs dark matter fraction (left). The gray regions mark the limitations of our analysis, \vtwo{where the EDO's radius is of order of its corresponding Schwarzschild radius}.}
    \label{fig:contourtogether}
\end{figure}

We present our results in Figure~\ref{fig:contourtogether}. On the plot on the left, {we express the EDO radius which leads to large $\Delta x_e$ as a function of the mass,} while keeping $f_{\rm DM}$ constant. We can see that the contour for $f_{\rm DM}=1$ perfectly agrees with the result from Ref.~\cite{Bai:2020jfm}, while we extend the calculations also to present constraints for $f_{\rm DM}=\{10^{-1},10^{-2},10^{-3}\}$, finding a logarithmic shift of the contours to higher masses for the same radii. On the right, we keep the radius constant, constraining EDOs with a given mass constituting some fraction of the total dark matter, going down to $f_{\rm DM}=10^{-5}$. These results complete the impact that extended dark objects may have on the CMB anisotropies, which can now be added to the rest of constraints via different tests, as will be discussed in the next section.

 \begin{figure}
    \centering
    \includegraphics[width=\textwidth]{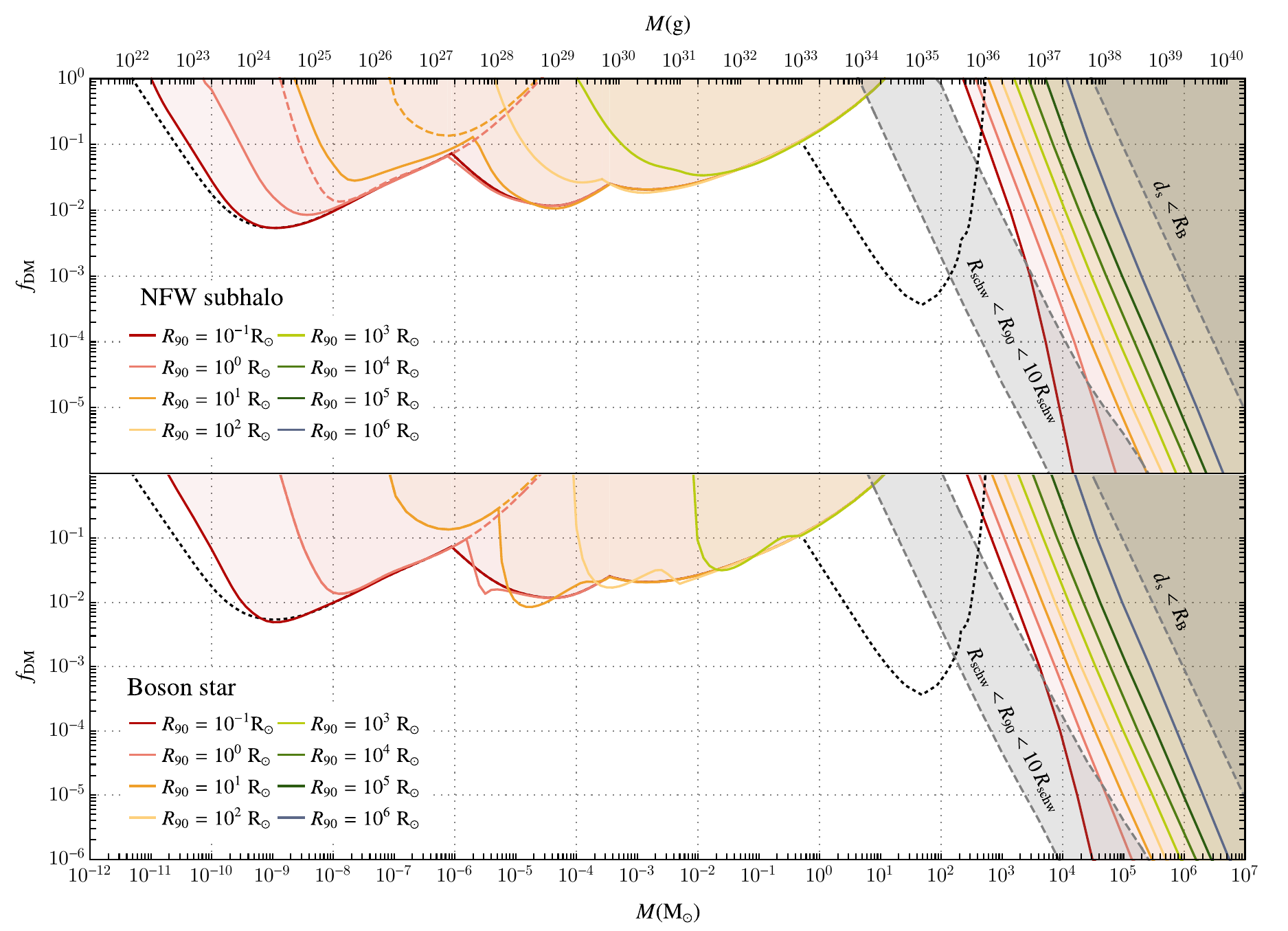}
    \caption{Complementarity of constraints on the maximum allowed dark matter fraction as a function of mass, for boson stars and NFW subhaloes of particular sizes. Shown here are constraints from the non-observation of microlensing signals \cite{Croon:2020ouk,Croon:2020wpr} in the EROS-2~\cite{EROSMACHOCombined}, OGLE-IV~\cite{Niikura:2019kqi} (note that this does not include the latest results based on a combination of the OGLE-III and OGLE-IV datasets \cite{Mroz:2024mse}), and Subaru-HSC~\cite{Subaru}  datasets;
    gravitational wave signals from the inspiral regime of binary mergers assuming binary formation in the early Universe (\cite{Croon:2022tmr}), and collisional ionisation of the CMB due to accretion (this work). {The gray regions show the limitations of our analysis, \vtwo{corresponding to those EDOs whose radius is comparable to their corresponding Schwarzschild radius, or those whose Bondi radius breaks the isolated EDO assumption (Eq.~\eqref{eq single EDO assumption})}.}}
    \label{fig:BScomplementarity}
\end{figure}
\section{Discussion}
Extended dark matter objects {such as boson stars, subhaloes and miniclusters} are {popular} dark matter candidates, {associated with a variety of creation mechanisms and potential mass ranges.} So far, {gravitational microlensing and gravitational waves from binary mergers have been used to}
constrain a wide range of the EDO parameter space {$ \lesssim 10 \, M_\odot$}. At large masses, {the ionisation of the} CMB 
due to the accretion of baryonic matter, which for PBHs {implies the leading constraint, can be used to constrain the dark matter fraction of these extended objects.} In Ref.~\cite{Bai:2020jfm}, the authors calculated the bounds for EDOs with uniform-sphere mass functions making up the $100\%$ of dark matter. In this work we extend this analysis in two important ways: we study different mass functions, and allow for these objects to be a subcomponent of DM. 
Specifically, besides uniform spheres \eqref{eq:uniformsphere} we focus on mini-haloes \eqref{eq:ucmh}, boson stars, and NFW subhaloes \eqref{eq:nfw}.
This allows
us to place large-mass constraints on EDOs of {these classes. We explain how our analysis can be easily extended to account for other dark objects. }

{The impact of different mass profiles is rather subtle, and primarily affects the interpretation of constraints.}
As described in Figure~\ref{fig: Luminosity}, {in the scenarios studied in this work} changing the mass function is equivalent to rescaling the radius for the uniform sphere. However, the rescaling of the constraints {between mass functions}
is model-dependent and non-trivial, so one first needs to calculate the averaged luminosity for each case to find the ratio between them. Once this is obtained, the constraints from Figure~\ref{fig:contourtogether} can be mapped into the new rescaled radii. To ease up this process, we provide the Mathematica notebook used to numerically generate the constraints for the uniform sphere,\footnote{The repository can be found in the following link: \href{https://gitlab.com/SergioSevillano/edo-accretion}{https://gitlab.com/SergioSevillano/edo-accretion}.} which works on any mass function as long as it can be expressed as $M(r/R)$ {where $R$ is defined such that $M(1)=M_{\rm tot}$} (see Appendix~\ref{App code}).

{The constraints we find are complementary to those found through the non-observation of gravitational microlensing events and binary mergers.}
In Figure \ref{fig:BScomplementarity} we show these complementary constraints for boson stars and NFW subhaloes together with
the CMB constraints derived in this work. As the gravitational wave constraints were derived assuming the inspiral regime, in which the waveform of these object binaries would be known, the objects in the plot have too low of a compactness ($C\lesssim 10^{-3}$) to be observed by the LVK collaboration, as the inspiral regime would have ended at frequencies outside of its observational window. {The apparent open window for stellar mass objects with stellar radius can be closed through gravitational heating of dwarf galaxies \cite{Graham:2023unf,Graham:2024hah}.} \vtwo{There is additionally a wide range of already existing PBHs constraints that may similarly apply to EDOs, such as radio and X-ray emission~\cite{Manshanden:2018tze}.}

In summary, this study extends the exploration of DM by introducing constraints on large mass EDOs, setting upper limits on their allowed mass fraction for a large range of masses and sizes. We have worked out constraints on boson stars, NFW subhaloes, and miniclusters in detail, and provided a Mathematica package to generalise our results to arbitrary other EDO profiles. Our results complement existing constraints derived from the non-observation of gravitational microlensing, binary mergers, and dynamical heating of dwarf galaxies. 

\section*{Acknowledgements}
We thank Joachim Kopp, Vivian Poulin, Harikrishnan Ramani, and Jens Chluba for useful discussions. DC and SSM are supported by the STFC under Grant No.~ST/T001011/1.

\appendix
\section{Improving the efficiency of the integral over relative velocities}\label{App code}
In section~\ref{sec:accretion} we showed how to obtain the temperature, density, and ionisation  fraction for the accreted matter around an EDO, and in section~\ref{sec: ionisation } the impact that they have on recombination history. Allowing for generic mass functions requires solving this system numerically, which is computationally expensive, especially when averaging the Luminosity for relative speeds of the EDOS. This requires performing the Gaussian integral over $v_{\rm rel}$ from Eq.~\eqref{eq: gaussian vrel},
\begin{equation}\label{eq: gaussian vrelApp}
    \langle\mathcal{L}\rangle=\frac{4\pi}{(2\pi\langle v_s^2\rangle/3)^{3/2}}\int_0^\infty \d v_{\rm rel}\, v_{\rm rel}^2 e^{-\frac{v_{\rm rel}^2}{2\langle v_s^2\rangle/3}}\mathcal{L}|_{c_\infty\to\sqrt{c_\infty^2+v_{\rm rel}^2}},
\end{equation}
where, a priori, for every evaluated value of the integrated, we would need to follow all the steps from section~\ref{sec:accretion} with a shift in the sound speed at infinity by $c_{\infty}^2\to c_{\infty}^2+v_{\rm rel}^2$.  However, in this appendix, we will show how one can \textit{recycle} EDO solutions for this problem so that this integral can be solved more efficiently.

This shift in the sound speed leads to a shift of the Bondi radius, $R_{\rm B}\to GM/(c_{\infty}^2+v_{\rm rel}^2)$, which only impacts the density equation,
\begin{equation}\label{eq: Main DE App}
	\frac{GM(r)}{r^2}+\gamma K \rho(r)^{\gamma-2} \frac{d \rho(r)}{dr}=0,
\end{equation}
through the constant factor $K$, defined outside of the ionizing region as
\begin{equation}
    K=\frac{GM}{\gamma_\infty R_{\rm B} \rho_\infty^{\gamma_\infty-1}}.
\end{equation}
\begin{figure}
    \centering
    \includegraphics[scale=0.5]{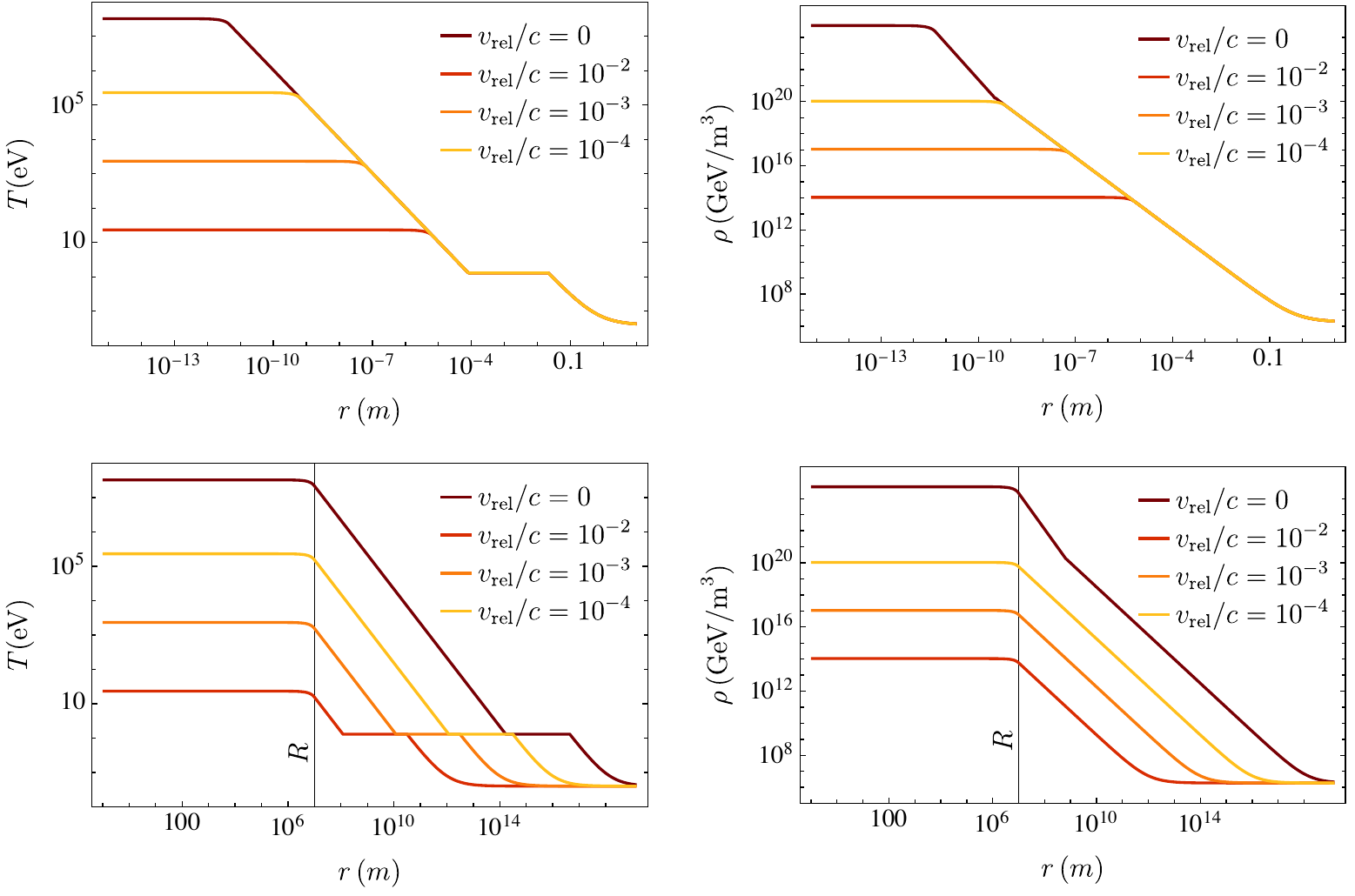}
    \caption{Example for the rescaling technique to save computing time when evalutating different relative velocities. The top two plots show the system with a rescaled radial component, $r'=r R_{\rm B}$, while the plots on the bottom correspond to the real solutions, obtained by inversing the rescaling for $r'$.}
    \label{fig:appendix}
\end{figure}
Therefore, in what follows, we will show that Eq.~\ref{eq: Main DE App} is independent of the Bondi radius by conveniently rescaling most quantities. For this, we will have to split the calculation into two regions
\begin{itemize}
    \item \textbf{Outside the EDO ($r>R$):} This region is defined as where the mass function becomes a constant ($M(r>R)=M)$. Thus, inserting $K$ into Eq.~\eqref{eq: Main DE App} and rescaling  $r\to R_{\rm B}r'$ and $\rho(r)\to \rho_\infty\rho'(r')$ gives
    \begin{equation}\label{eq: Out DE App}
	\frac{1}{r'^2}+\rho'(r')^{\gamma-2} \frac{d \rho'(r')}{dr'}=0,
\end{equation}
which can be solved for $r'\leq1$, and matched at the boundary ($r'=R/R_{\rm B}$) with the interior solution.
    \item \textbf{Inside the EDO ($r\leq R$):} In this case, we will need to take into account the radial dependence of the mass function. The rescaling is possible as long as the mass function can be expressed as a function of $r/R$, which is generally the case. In this way, rescaling $r\to\hat{r} R$, $\rho(r)\to\rho_\infty\hat{\rho}(\hat{r})R^{\frac{1}{\gamma_\infty-1}}$ and $M(\hat{r})\to\tilde{M}(\hat{r})M_{\rm tot}$, we obtain
    \begin{equation}\label{eq: Main DE App}
	\frac{\tilde{M}(\hat{r})}{\hat{r}^2}+\hat{\rho}(\hat{r} )^{\gamma-2} \frac{d \hat{\rho}(\hat{r})}{d\hat{r}}=0,
\end{equation}
 which again does not explicitly depend on $R_{\rm B}$.
\end{itemize}
Therefore, the idea is to solve these equations without specifying $R_{\rm B}$, and then rescale and match the solutions at the boundary for each evaluated $v_{\rm rel}$ in the integral. Moreover, since the temperature of the accreted matter and its density are related via
\begin{equation}
    T(r)=T_\infty \left(\frac{\rho(r)}{\rho_\infty}\right)^{\gamma-1},
\end{equation}
and this does not depend on $R_{\rm B}$ either, we find that the rescaled quantities corresponding to the ionisation  start and end, $r'_{\rm start}=r_{\rm start}/R_{\rm B}$ and $r'_{\rm end}=r_{\rm end}/R_{\rm B}$, respectively, will not depend on $R_{\rm B}$ either, so they can also be rescaled after being computed for the general case. However, given that the relation between $T(r)$ and $\rho(r)$ depends on their values at infinity, we will have to compute the whole system for different redshifts.

In Figure~\ref{fig:appendix} we show an example of this rescaling for different relative velocities (upper plots); by rescaling $r\to r/R_{\rm B}$, the solutions perfectly agree up to the beginning of the different interior regions. However, this is only because in the rescaled units the boundary to the interior region is $R/R_{\rm B}$, which depends on the relative velocity. All lines correspond to EDOs with the same mass and radii, as can be seen in the un-rescaled plots at the bottom.  

\bibliographystyle{apsrev4-1}
\bibliography{main}
\end{document}